\documentclass[aps,prb,superscriptaddress,
 sd,%
 amsmath,amssymb,
 reprint,%
]{revtex4-1}

\pdfoutput=1

\usepackage{graphicx}
\usepackage{dcolumn}
\usepackage{bm}

\usepackage{color}

\begin{document}

\title{Large spin-mixing conductance in highly Bi-doped Cu thin films}%

\author{Sandra Ruiz-G\'omez}
\affiliation{Dept. F\'isica de Materiales. Universidad Complutense de Madrid. 28040 Madrid, Spain}

\author{A\'ida Serrano}
\affiliation{SpLine, Spanish CRG BM25 Beamline, ESRF, 38000 Grenoble, France}

\author{Rub\'en Guerrero}
\affiliation{Instituto Madrile\~no de Estudios Avanzados - IMDEA Nanociencia, 28049, Madrid, Spain}

\author{Manuel Mu\~noz}
\affiliation{Instituto de Micro y Nanotecnolog\'ia (CNM-CSIC), PTM, 28760 Tres Cantos, Madrid, Spain}

\author{Irene Lucas}
\affiliation{Dpto. F\'isica de la Materia Condensada, Universidad de Zaragoza, Pedro Cerbuna 12, 50009 Zaragoza, Spain}
\affiliation{Instituto de Nanociencia de Arag\'on (INA), Universidad de Zaragoza, Mariano Esquillor, Edificio I+D, 50018 Zaragoza, Spain}

\author{Michael Foerster}
\author{Lucia Aballe}
\affiliation{Alba Synchrotron Light Facility, CELLS, E-08290, Carrer de la Llum 2-23, Bellaterra, Spain}

\author{Jos\'e F. Marco}
\affiliation{Instituto de Qu\'imica F\'isica Rocasolano - CSIC, Calle de Serrano, 119, 28006 Madrid, Spain}
\affiliation{Unidad Asociada IQFR (CSIC)-UCM, 28040, Madrid, Spain}

\author{Mario Amado}
\author{Lauren McKenzie-Sell }
\author{Angelo di Bernardo}
\author{Jason W. A. Robinson}
\affiliation{Department of Materials Science and Metallurgy, University of Cambridge, 27 Charles Babbage Road, Cambridge CB3 0FS}

\author{Miguel \'Angel Gonz\'alez Barrio}
\affiliation{Dept. F\'isica de Materiales. Universidad Complutense de Madrid. 28040 Madrid, Spain}
\affiliation{Unidad Asociada IQFR (CSIC)-UCM, 28040, Madrid, Spain}

\author{Arantzazu Mascaraque}
\affiliation{Dept. F\'isica de Materiales. Universidad Complutense de Madrid. 28040 Madrid, Spain}
\affiliation{Unidad Asociada IQFR (CSIC)-UCM, 28040, Madrid, Spain}

\author{Lucas P\'erez}

\affiliation{Dept. F\'isica de Materiales. Universidad Complutense de Madrid. 28040 Madrid, Spain}
\affiliation{Instituto Madrile\~no de Estudios Avanzados - IMDEA Nanociencia, 28049, Madrid, Spain}
\affiliation{Unidad Asociada IQFR (CSIC)-UCM, 28040, Madrid, Spain}

\email{lucas.perez@fis.ucm.es}

 \begin{abstract}
  {\bf Abstract}
  
  Spin Hall effect provides an efficient tool for the conversion of a charge current into a spin current, opening the possibility of producing pure spin currents in non-magnetic materials for the next generation of spintronics devices. In this sense, giant Spin Hall Effect has been recently reported in Cu doped with 0.5\% Bi grown by sputtering and larger values are expected for larger Bi doping, according to first principles calculations. In this work we demonstrate the possibility of doping Cu with up to 10\% of Bi atoms without evidences of Bi surface segregation or cluster formation, as studied by different microscopic and spectroscopic techniques. In addition, YIG/BiCu structures have been grown, showing a spin mixing conductance larger that the one shown by similar Pt/YIG structures. These results reflects the potentiality of these new materials in spintronics devices. 
    \end{abstract}

\maketitle
\renewcommand{\tablename}{Table}

\section{Introduction}

Spintronics studies and exploits the intrinsic spin of the electron and its associated magnetic moment jointly with its fundamental charge for the generation, manipulation, and detection of  spin currents that can be implemented in solid state devices\cite{zutic04}. In particular, the possibility of harvesting pure spin currents, i.e., without charge current,  and use them in memory-logic devices plays a key role in the next generation of electronics due to the expected low power consumption \cite{favio12,chumak15}. In this regards, spin Hall effect (SHE) provides an efficient tool for the conversion of a charge current into a spin current in nonmagnetic materials without any associated current injection from the ferromagnets\cite{jungwirth12,Sinova15}. To exploit this effect, identifying novel materials with a large charge-to-spin conversion, i.e. materials with large spin Hall angles (SHA) turns out to be essential \cite{wang17,jaya14,jin17}.

In addition to metals showing intrinsic SHE, giant SHE has been predicted when they are conveniently doped with impurities showing strong spin-orbit interactions\cite{Gradhand10,niimi15}. In these materials, extrinsic scattering  mechanisms such as skew scattering\cite{herschbach13} and side jump effect \cite{fert11} might also give rise to SHE. In fact,a SHA of $\sim~-0.24$ has been measured by Niimi and coworkers in CuBi alloys with $\sim~0.5$\% of Bi grown by sputtering \cite{niimi12}. The authors mention in the Supplemental Material of reference \onlinecite{niimi12} that a larger Bi content leads to surface segregation. However, larger values of SHA are expected theoretically for higher doping of Bi in Cu\cite{fedorov13}.

In this work we demonstrate Bi doping of Cu films with up to 10\% without surface segregation or cluster formation. When grown on yttrium iron garnet (YIG) substrates, the interfaces show a large spin mixing conductance. These new materials open the possibility of studying the giant SHE and using it in future spintronics devices.

\section{Experimental section}

The first growth stages were studied at the CIRCE beamline\cite{aballe15} of the Alba synchrotron in Spain. The beamline experimental station chamber houses an Elmitec III Low-Energy Electron Microscope  (LEEM) that allows for fast real-space imaging of the surface of the CuBi films during growth, as well as selected-area low energy electron diffraction measurements (LEED). Samples for the LEEM experiments were grown in-situ onto Ru(0001) single-crystal substrates that were cleaned by exposure to $10^{-8}$~mbar of oxygen at 1000K followed by flashing at 1500 K in vacuum. Afterwards, Bi and Cu were co-evaporated from electron-beam heated dosers, at a base pressure of $1\times10^{-10}$~mbar. The Bi and Cu evaporation rates were adjusted to ensure that the Bi content in the film was below 10\% in all experiments. 

Thin films were grown by thermal evaporation in a High-Vacuum (HV) chamber with individual Joule dosers for Bi and Cu. The base pressure of the chamber was $10^{-7}$~mbar. Previously to the films growth, the dosers were calibrated using a quartz crystal microbalance. The substrates were Si(100) wafers with native SiO$_2$ and were kept at room temperature during growth. 

A scanning electron microscope (SEM) JEOL JEM 6335 equipped with an energy dispersive X-ray (EDX) system was used to study the composition of the samples. The crystalline structure was studied by X-ray diffraction (XRD) with a grazing incidence  angle of 0.5$^\circ$, with a PANalytical X-ray diffractometer using CuK$_\alpha$ radiation. Scanning transmission electron microscopy (STEM) was performed in an FEI Titan 60 operated at 300 kV and equipped with a high brightness Schottky field emission gun. Z contrast imaging has been carried out in high angle annular dark field (HAADF) with a probe convergence angle of 25 mrad and an inner collection angle of 58 mrad. The specimens analyzed are lamellas extracted from the samples by focused ion beam (FIB) milling in a FEI Helios Nanolab 600.

X-ray Photoelectron Spectroscopy (XPS) data were recorded with a CLAM2 analyser under a base pressure of $5\times10^{-9}$~mbar using Mg K$\alpha$ radiation and a constant pass energy of 200 eV and 20 eV for the wide and narrow scan spectra, respectively. The binding energy scale was referenced to that of the C 1s signal of the adventitious contamination layer which was set at 284.6 eV. The samples were subjected to Ar$^+$ sputtering to investigate the in-depth distribution of the different chemical elements. An integral Ar$^+$ ion gun was used at 20 keV and 20 mA and an Ar pressure within the analysis chamber of $5\times10^{-5}$~mbar. 

X-ray absorption spectroscopy (XAS) measurements is carried out at the Cu K-edge (8.98~keV) and Bi L$_3$-edge (13.42~keV) at the beamline BM25A of The European Synchrotron (ESRF) in Grenoble (France).  Bi and Cu foils were measured in transmission mode at the beginning of the experiment for energy calibration. Cu$_2$O, CuO and Bi$_2$O$_3$ reference samples were also measured in transmission mode. Samples spectra were collected in fluorescence yield mode placing samples at 45$^\circ$ from the incoming X-ray beam and forming 45$^\circ$ with the dispersive X-ray fluorescence detector. From three to five spectra were acquired from each sample and merged in order to improve the signal-to-noise ratio.  The acquisition time for each energy scan was about 40 min. XAS data were processed using the Demeter package and applying standard methods\cite{Ravel05}.

Ferromagnetic resonance spectroscopy (FMR) was carried out at room temperature (RT) on BiCu/YIG (110) structures prepared as described in the Supporting Information. The samples were placed face down in a grounded coplanar waveguide in a wide band set up in order to excite the sample with the RF field. The DC external magnetic field was generated by an electromagnet, and aligned perpendicular to the RF field and parallel to the sample (in-plane geometry). Microwave signal (constant power of -5 dBm) was applied to the waveguide using a KEYSIGHT N9918A generator, and a diode KEYSIGHT 8473B was used for detection. Field modulation (0.6 Oe) and a lock-in amplifier were used to extract the derivative of the absorbed power versus DC field. 

\section{Results and discussion}

\begin{figure}
\centering{\includegraphics{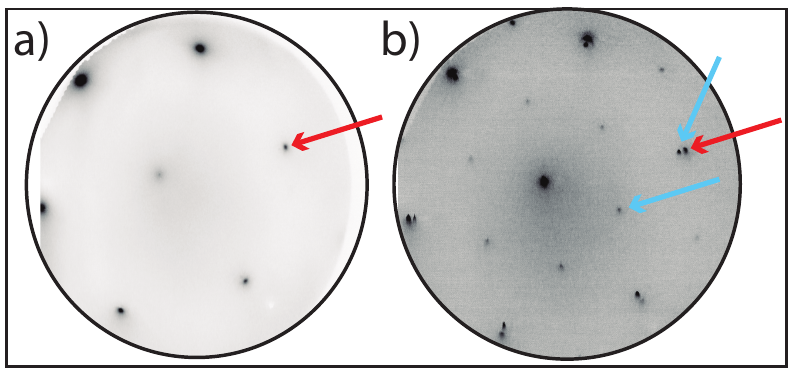}}
\caption{\label{figure1} (a) LEED pattern of a Cu(Bi) thin film grown on Ru(0001) at RT. (b) LEED pattern of a Cu(Bi) thin film grown on Ru(0001) at $100^\circ$C. The red arrows mark characteristics spots of Cu(111) on Ru(0001) and the blue arrows spots of the ($\sqrt{3}\times\sqrt{3}$) R30  reconstruction of Bi  on Cu(111).}
\end{figure}

Bi has been widely used as surfactant in the growth of metals in UHV vacuum conditions \cite{young15}. Therefore, as a first step it is important to establish the growth conditions to avoid Bi migration to the film surface. We have therefore studied the first stages of growth of the model system Cu/Ru(0001) at different temperatures, incorporating Bi in  Cu during growth while monitoring the process using LEED pattern.  Figure~\ref{figure1}.a shows the LEED pattern obtained during the growth at room-temperature. The spots observed in the LEED pattern correspond to the hexagonal ones of the Cu(111) on Ru (marked with a red arrow). There are no extra-spots corresponding to Bi on a Cu surface, i.e., the Bi atoms are either not on the Cu surface or they do not form an ordered structure. However, when Cu and Bi are co-evaporated at a higher temperature ($100^\circ$C) (Figure~\ref{figure1}.b) additional LEED spots appear that can be related to a ($\sqrt{3}\times\sqrt{3}$) $R30^\circ$ reconstruction, which is characteristics of a Bi monolayer on the surface of a Cu(111)\cite{gastel14}. From the LEED pattern, it is clear that, when growing at high temperature, Bi segregates to the surface whereas no segregation is observed at RT.

If these Bi-doped Cu films are intended to be used in spintronics, it is mandatory to grow them on substrates that allow for the realization of magnetotransport measurements. Taking this into account, samples were grown on SiO$_2$/Si at RT.  The Cu evaporation rate is kept constant at 0.4~\AA/s while the Bi evaporation rate was varied to obtain a Bi content in Cu from 1\% up to 40\% (wt.) --- as measured by EDX. Pure Cu and Bi films were also grown as reference samples. EDX measurements reveal an homogeneous composition in all samples.

A first approach of the Bi distribution along the cross section of the BiCu films can be obtained from XPS technique. Figure~S1 collects the wide scan spectra recorded at different sputtering times from a Bi$_{85}$Cu$_{15}$ sample. The spectra show only Cu, Bi, O and C signals. As the sputtering time increases, the intensity of the C and O signals decreases strongly while the Cu 2p peaks increases, due to removal of the contamination layer from the uppermost surface being really small after long sputtering times. The spectra also show a clear increase of the intensity of the Cu 2p peaks with increasing sputtering time. However, more than due to an increase in Cu concentration this must be related to a much smaller, almost insignificant, attenuation of the Cu 2p electrons once the contamination layer has been removed from the uppermost surface (see below). 

\begin{figure}
\centering{\includegraphics[width=1\columnwidth]{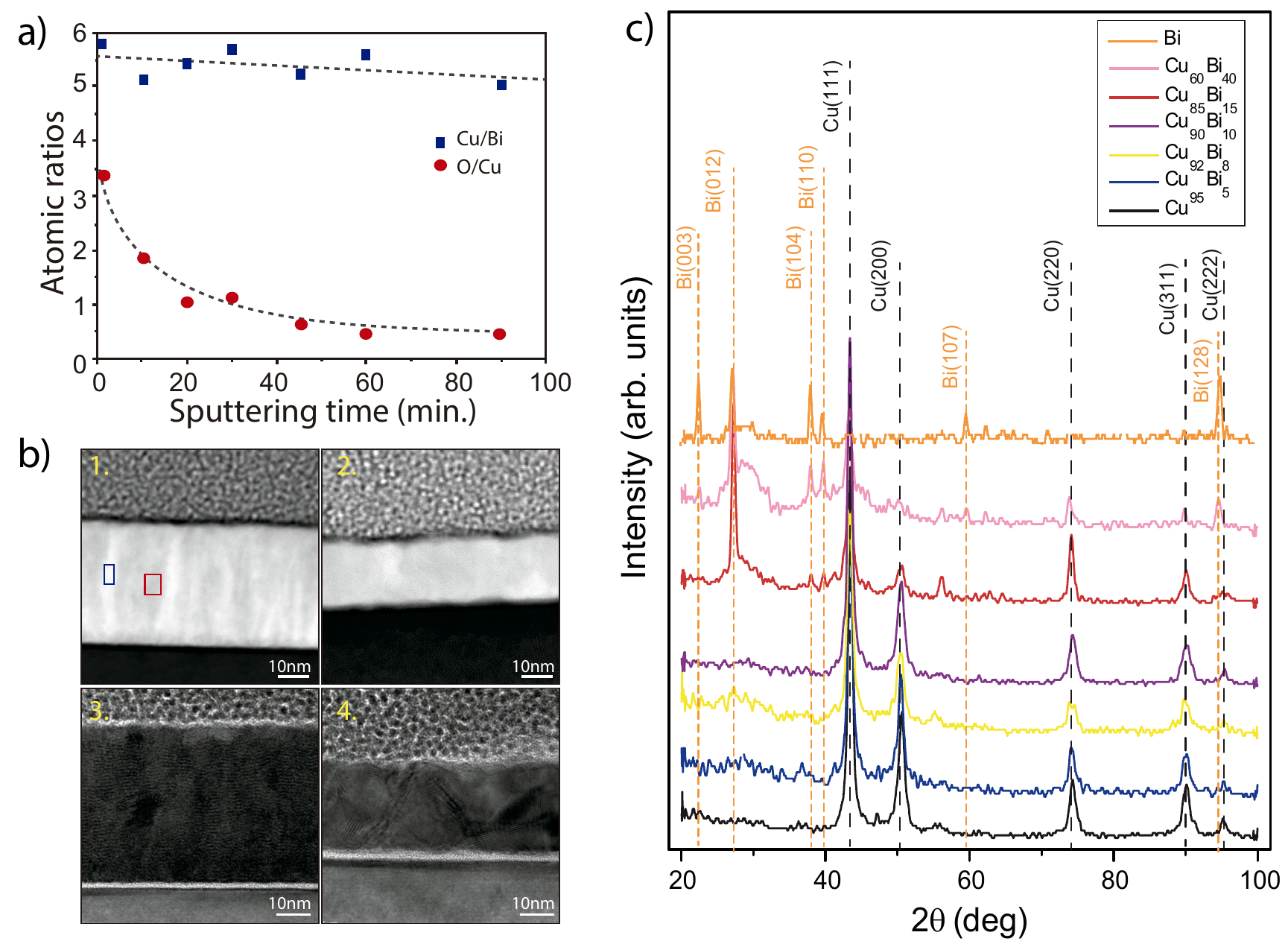}}
\caption{\label{figure3} (a) Atomic ratios obtained from the evaluation of the XPS data for a Cu$_{75}$Bi$_{15}$ film. (b) (1) Z-contrast and (3) HR-TEM image of Cu$_{75}$Bi$_{25}$, (2) Z-contrast and (4) HR-TEM image of Cu$_{96}$Bi$_{4}$. (c) XRD patterns for the Bi-doped samples under study. Reflections corresponding to the Cu-fcc spectrum are marked with gray dashed lines and the ones corresponding to the orthorhombic Bi structure with orange dashed lines.}
\label{Fig2}
\end{figure}

Cu/Bi and O/Cu atomic ratios were calculated from the integration of the Cu 2p, O 1s and Bi 4f spectral areas after background subtraction (Shirley method) using the Multiquant XPS software. This package allows taking into account the attenuation brought about by the adventitious carbon layer\cite{mohai04}. This is important in the present case, particularly for the Cu/Bi atomic ratio, since the Cu 2p and Bi 4f spectral regions are separated by approximately 800 eV and therefore the signal corresponding to the Cu 2p electrons, that have a much smaller kinetic energy than the Bi 4f ones, is significantly more attenuated by the contamination layer than that of the Bi 4f electrons. Figure \ref{Fig2}.a  shows the Cu/Bi and O/Cu atomic ratios obtained from the XPS data. It is clear from Figure 2 that the O/Cu ratio decreases rapidly with sputtering time, indicating that the surface oxidic layer is effectively removed by Ar$^+$ bombardment  (XPS data, not shown, reveal that the most external part of the sample contains both Cu$^{+}$ and Bi$^{3+}$). Contrarily, the Cu/Bi ratio remains fairly constant with sputtering time both within the oxidic surface layer and in the sample itself. Given the composition of the sample, the expected Cu/Bi atomic ratio should be 5.7 which is close, within the error of the experimental determination, to the value found in the current experiments. Therefore, in view of the present data, we can conclude that there is no enrichment either in Cu or Bi within the depth explored in this work.

Additional information on the distribution of Bi in the alloys can be extracted from cross-sectional high resolution TEM images. Panel 1 of Figure~\ref{Fig2}.b shows a TEM image with Z-contrast measured in  the sample Cu$_{75}$Bi$_{25}$. Dark and light areas can be clearly distinguished, corresponding to zones with different composition. In particular, blue square corresponds to a Bi-poor region (light area) whereas red square to a Bi-rich region (dark area). From this image it is noted a clear segregation of Bi in the sample. However, the Z-contrast image measured in the sample Cu$_{96}$Bi$_{4}$  (Panel 2 of Figure~\ref{Fig2}.b) is much more homogeneous, without a clear contrast. In this case Bi is distributed homogeneously across the entire sample. No Bi accumulation has been observed in either top or bottom surfaces of the thin films. The high-resolution images (Panels 3 and 4 in Figure~\ref{Fig2}.b) evidence the polycrystalline nature of the samples where the disorder increased with Bi content.

\begin{figure*}
\centering{\includegraphics[width=1\textwidth]{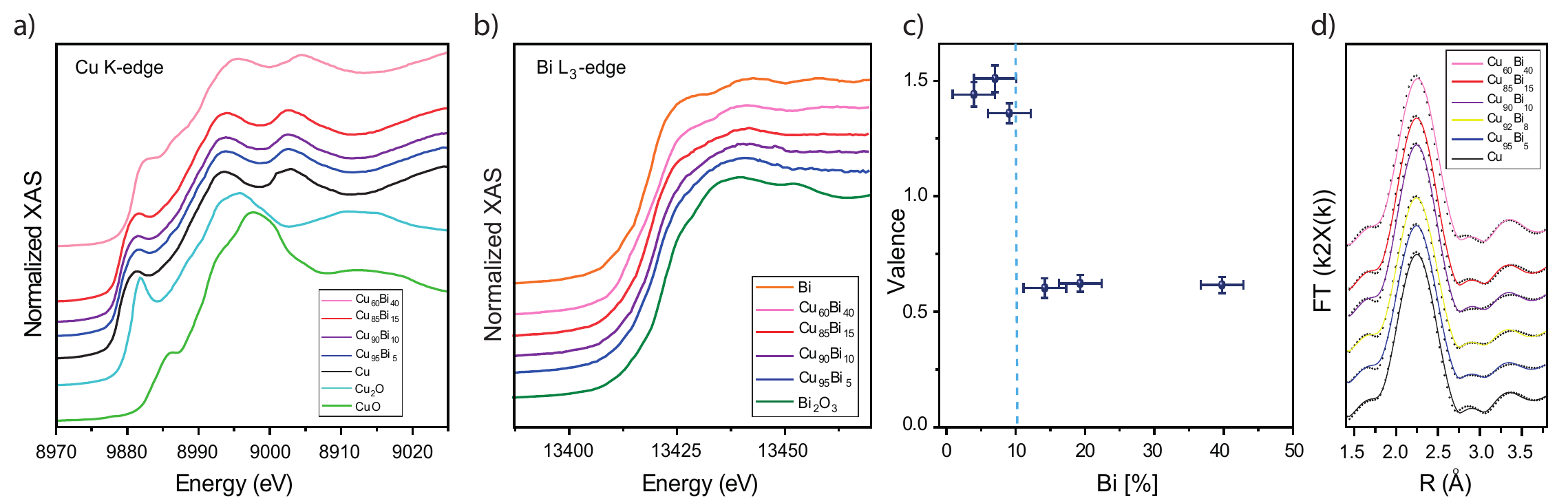}}
\caption{\label{Fig3} (a) Normalized XANES spectra at Cu K-edge of the different  Cu$_\text{1-x}$Bi$_\text{x}$ thin films together with the Cu, CuO and CuO reference samples.(b) Normalized XANES spectra at the Bi L3-edge of  Cu$_\text{1-x}$Bi$_\text{x}$ thin films and Bi and Bi$_2$O$_3$ references.(c) valence of the Bi atoms as a function of the Bi content of the alloys, calculated as described in the text. (d) Fourier transform function of EXAFS signal (experimental and fitting) of films from signal collected at Cu K-edge.}
\label{FIG3}
\end{figure*}

The presence of large clusters can be explored by XRD. Considering the thickness of the films (below 50~nm in all cases), we have measured in grazing incidence configuration to reduce the signal from the substrate and get a better signal-to-noise ratio. Figure~\ref{Fig2}.c collects the XRD patterns for the thin films under study, together with a Cu and Bi reference thin film. Reflections corresponding to the Cu-fcc structure are marked with black lines, the ones corresponding to the orthorhombic Bi structure with orange lines and the one marked with an asterisk correspond to the Si substrate. As expected, Cu and Bi reference samples only show reflections that can be indexed as pure Cu or Bi respectively. For low doped samples (Cu$_{95}$Bi$_{5}$, Cu$_{92}$Bi$_{8}$ and Cu$_{90}$Bi$_{10}$), we observe Cu(111), Cu(200) and Cu(220) reflections, the same as in the Cu reference. There is only a slight increase in the width of the peaks, probably due to the disorder induced in the Cu lattice by the incorporation of Bi atoms. No Bi reflections are detected. The Bi atoms seem incorporated into the fcc-Cu lattice. When increasing the doping level above 10\%, see samples Cu$_{85}$Bi$_{15}$ and Cu$_{60}$Bi$_{40}$, new reflections appear that can be indexed as Bi(012) and Bi(311). Thus, as expected, for high Bi doping Bi agglomerates and clusters are detected by XRD.

To further investigate the structure of the films and to elucidate whether Bi clusters are formed into the Cu matrix, we employed X-ray absorption fine structure techniques, which are sensitive to the local atomic environment of X-ray absorbing atoms. Figure~\ref{Fig3}.a displays the normalized X-ray absorption near edge structure (XANES) spectra at the Cu K-edge for the different  Cu$_\text{x}$Bi$_\text{y}$ thin films together with the Cu foil, Cu$_2$O and CuO references. A close inspection of the different curves reveals that the incorporation of Bi does not change the shape of the spectra or the position of the energy edge unless the Bi content exceeds 15\%. In the Cu$_{60}$Bi$_{40}$ sample, a shift in the absorption edge and the resonances after the edge towards higher energies, as well as dissimilarities in the line shape, are observed. This might be a clear indication of the introduction of a large disorder in the Cu structure. To quantify the possible oxidation of the layers, the XANES spectra of the samples were fitted to a linear combination of the spectra of  different reference samples. All samples show an oxide content around 8\%, independent of the Bi doping. This oxide could be ascribed to a surface oxidation layer, as supported by XPS data shown before. Visual examination of the XANES spectra measured at Bi L$_3$-edge (Figure~\ref{Fig3}.b) reinforces the idea of Bi diluted in the Cu matrix. Spectra corresponding to the samples with Bi content below 15\% are very different from the Bi reference sample as well as from the Bi$_2$O$_3$ reference sample, which reflects a different local environment of the doping Bi atoms.

The mean valence of the Bi doping atoms in the structure can be calculated using Kunzl's law \cite{Kunz}, by linear interpolation of  the shift of the edge position with respect to the absorption edge of Bi metallic and oxide references. The output of this interpolation as a function of the Bi content in the thin films is shown in Figure~\ref{Fig3}.c. It is possible to see that, within the noise, there are only two set of values for the valence related to Bi in Cu matrix and Bi in clusters form, with a limiting value around 10\%. It is particularly noticeable that the value of the valence obtained for samples with a content of bismuth below ~10\%, a value close to 1.5, is in good agreement with reported values obtained form ab initio calculations in homogeneously Bi-doped Cu films \cite{levy13}. 

\begin{table}
\small
  \caption{Nearest neighbour structural parameters obtained by the FT $k^2 \chi(k)$ curve fitting for Cu foil reference and Cu(Bi) films. N is the coordination number that has fixed for the Cu foil reference, R$_\text{Cu-Cu}$ is the average interatomic distance and $ \sigma^2$ are the Debye-Waller factors. }
  \label{tbl:Artemis}
 
  \begin{tabular*}{0.5\textwidth}{@{\extracolsep{\fill}}ccccccc}

    \hline
    
    \%Bi & N &	R$_\text{Cu-Cu}$(\AA)& $ \sigma^2$ ($ \times 10^{-3}$~\AA$^2$)\\
    \hline

Foil &	12 &	2.544 (2) & 8.76 (9) \\
0 &	11.1 (2) &	2.543 (2) & 8.6 (1)  \\
5 &	10.6 (1) & 	2.542 (7)	 & 8.7 (1)  \\
8 &	10.1 (1) &	2.542 (1) &  8.40 (9)  \\
10 &	10.8 (1) &	2.543 (9) &  8.7 (1)  \\
15 &	10.8 (1) & 	2.544 (3) &  8.6 (1)  \\
40 &	9.9 (2)   &	2.548 (5) &  9.0 (2) \\

    \hline
  \end{tabular*}
\end{table}

\begin{table*}

\small
  \caption{Damping constants and spin mixing conductance for bare YIG substrates as well as for YIG/CuBi interfaces.}
  \label{tbl:fmr}
 
  \begin{tabular*}{\textwidth}{@{\extracolsep{\fill}}ccccccc}

    \hline
    
    Sample & $\alpha_\text{YIG} (\times 10^{-3})$ & $\alpha_\text{YIG/CuBi} (\times 10^{-3})$ &	$\alpha_\text{sp} (\times 10^{-3})$ & $G_\text{eff} ( \times 10^{18} m^{-2})$\\
    \hline

Cu$_{99}$Bi$_1$ &	$3.7 \pm 0.2$ &	$5.4 \pm 0.2$ & $1.7\pm 0.4$ & $7.1 \pm 0.7 $ \\
Cu$_{96}$Bi$_4$ &	$1.7 \pm 0.1$ &	$3.7 \pm 0.1$ & $2.0\pm 0.2$ & $7.3 \pm 0.5 $\\

    \hline
  \end{tabular*}
\end{table*}

Analysis of the Fourier transform (FT) of the extended X-ray absorption fine structure (EXAFS) spectra was performed for several Bi-doped Cu films as well as the Cu foil reference. A $k^2$ weighting was used in the $k$ range $2.7-13.0$~\AA$^{-1}$ for fitting of FT signals in R space using theoretical functions from the FEFF code \cite{Ankudinov98}. Experimental FT module and the fitting is shown in Figure~\ref{Fig3}.d. Table~\ref{tbl:Artemis} displays the fitting structural parameters. FT $k^2 \chi(k)$  fitting of the first interatomic distance is very similar for all samples under study and to that of Cu foil. Structural parameters obtained at the first Cu-Cu shell do not alter at local order the Cu structure except for sample with a 40\% Bi in which the incorporation of  Bi atoms into the Cu lattice induces a slight decrease of coordination number, an elongation of Cu-Cu distance and an increase of the Debye-Waller factor. Fitting of second shell in each FT $k^2 \chi(k)$ spectrum was also performed. However these do not show a clear tendency with Bi content. It seems that for low Bi content, Bi atoms incorporates in the Cu lattice, producing a distortion of it at larger local order, which makes difficult to analyze the second shell considering slight distortions of Cu structure.

Finally, we performed FMR measurements on two YIG/BiCu heterostructures. The FMR was measured as described in the supplementary information. In order to calculate the damping constant, the measurements were performed at different values of the excitation frequency.

\begin{figure}
\centering{\includegraphics[width=1\columnwidth]{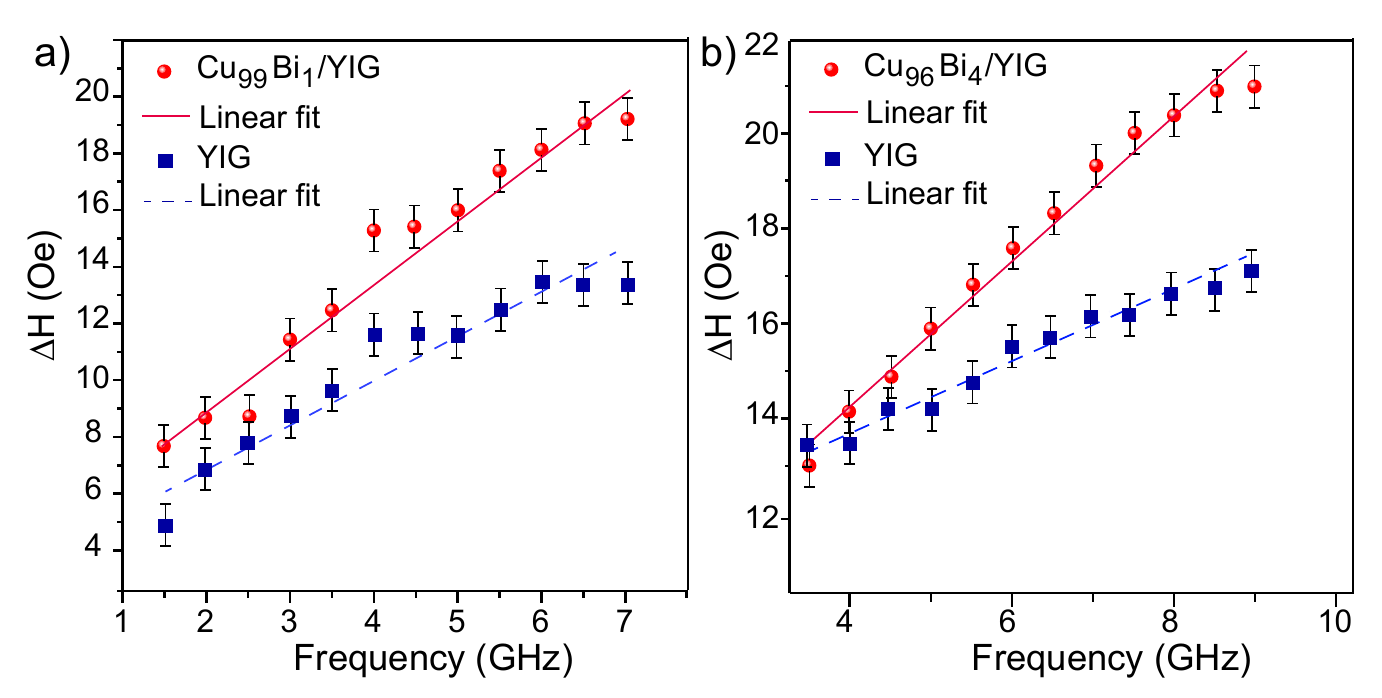}}
\caption{\label{figurefmr} (a) Frequency dependence of the FMR linewidth for sample Cu$_{99}$Bi$_1$ after and before CuBi deposition and (b) sample Cu$_{96}$Bi$_4$ after and before CuBi deposition.}
\label{Fig4}
\end{figure}

Using Kittel's equation for in plane measurements \cite{kittel48, hauser15} it is possible to obtain the values of effective saturation magnetization and the Land\'e gyromagnetic factor ($\it{g}$). The continuous line in Figure S3 corresponds to the fitting of experimental data using this equation

\begin{equation}
f=| \gamma | \sqrt{H_\text{FMR} (H_\text{FMR} + 4 \pi M_{s} - H_\text{ani})}
\end{equation} 

Regarding the YIG substrates whose measurements are shown in Figures S2, we have found a $4\pi M_S$ of 1730 G in the first case and of 1493 G in the second case. The value of the gyromagnetic ratio is 2.8 MHz/Oe in both cases, which is in good agreement with reported values \cite{haertinger15,hauser15,hahn13,hide11}. Plotting now the dependence of the linewidth with frequency, the damping can be calculated from the following equation: 

\begin{equation}
\Delta H_\text{FMR}=\Delta H_{0}+ \frac{2 \alpha f}{\sqrt{3} \gamma}
\end{equation} 

After capping the YIG with a CuBi layer, a significant increase in the slope of the frequency-dependent linewidth and hence an increased Gilbert damping constant is observed (See Figure \ref{figurefmr}). When a ferromagnetic layer as YIG is capped with a metallic layer as CuBi, the precession of the magnetization in the magnetic layer causes a flow of spins to the metallic layer because CuBi acts as a spin sink. Therefore, the damping constant for YIG/CuBi is the damping constant for uncapped YIG plus a contribution due to spin pumping $\alpha_\text{0}=\alpha + \alpha_\text{sp}$ . The damping obtained for both samples is summarized in Table~\ref{tbl:fmr}. Due to conservation of angular momentum, this additional damping can be used to evaluate the CuBi/YIG interface spin-mixing conductance. The additional Gilbert damping is related to the effective interface spin-mixing conductance $G_\text{eff}$ by the following relationship\cite{haertinger15}:

\begin{equation}
\alpha_\text{sp}= \frac{ g \mu _{B} 4 \pi M_S G_{eff}  }{t_\text{YIG}}
\end{equation}

where M$_{S}$ is the saturation magnetization and t$_\text{YIG}$ is the thickness of the magnetic material,  g is the g factor and $\mu_\text{B}$ is the Bohr magneton. The spin mixing conductance obtained for both samples is summarized in Table~\ref{tbl:fmr}. These values are in the same range than the values measured in optimized Pt/YIG interfaces\cite{Wang17b}

\section{Conclusions}

To sum up, we have demonstrated that it is possible to incorporate up to 10\% of Bi atoms into the Cu structure by co-evaporation of Bi and Cu atoms in a molecular beam epitaxy system at room temperature. Bi incorporates in the Cu lattice, without any trace of segregation or cluster formation below 10\% of Bi. There is also no presence of Cu or Bi oxides apart from the surface oxidation layer formed when the sample is exposed to air. Structural properties of Bi-doped Cu with up to 10\% Bi are similar to the one of Cu, reflecting the incorporation of Bi in the Cu structure, forming an alloy. CuBi/YIG interfaces have also been studied by FMR. These interfaces show a large spin-mixing conductance, which opens the possibility of exploring the spin hall effect of these alloys beyond the region explored up to date, expanding their possible use in spintronics.


\begin{acknowledgements}
This work has been partially funded by MAT2014-52477-C5, MAT2017-87072-C4 and MAT2015-64110-C2-2-P from the Ministerio de Ciencia e Innovaci\'{o}n and Nanofrontmag from Comunidad de Madrid. IMDEA Nanociencia acknowledges support from the Severo Ochoa Programme for Centres of Excellence in R\&D (MINECO, Grant SEV-2016-0686). We acknowledge The European Synchrotron Radiation Facility (ESRF), MINECO and CSIC for provision of synchrotron radiation facilities, BM25-SpLine staff for the technical support beyond their duties and the financial support for the beamline (PIE-2010-OE-013-200014).  We thank the Spanish National Center of Electron Microscopy for SEM measurements and the CAI de Difracci\'on de Rayos X, Universidad Complutense de Madrid, for XRD measurements. The TEM works have been conducted in the Laboratorio de Microscop\'ias Avanzadas (LMA) at the Instituto de Nanociencia de Arag\'on (INA)- Universidad de Zaragoza. Authors acknowledge the LMA-INA for offering access to their instruments and expertise. M.A. acknowledeges MSCA-IFEF-ST No. 656485-Spin3 J.W.A.R Acknowledges Royal Society (Superconducting Spintronics), Leverhulme Trust (IN-2013-033). 
\end{acknowledgements}

%
%
%

\nocite{*}
\bibliography{CuBi}

\end{document}